# Up-Converting Luminescent Nanoparticles as Probes of Surface Dynamics in Single Evaporating Microdroplets of Suspension

Yaroslav Shopa [a], Maciej Kolwas [b], Daniel Jakubczyk [b, *], Gennadiy Derkachov [b], Izabela Kamińska [b], Krzysztof Fronc [b, c], Tomasz Wojciechowski [b, c]

[a] *Cardinal Stefan Wyszynski University in Warsaw, Dewajtis 5 , 01-815 Warsaw, Poland*

[b] *Institute of Physics, Polish Academy of Sciences, al. Lotników 32/46, 02-668 Warsaw, Poland*

[c] *International Research Centre MagTop, al. Lotników 32/46, 02-668 Warsaw, Poland*

[*]Corresponding author. E-mail address: jakub@ifpan.edu.pl

## Abstract

We have investigated the optically measurable properties of single evaporating microdroplets of suspensions containing up-converting luminescent nanoparticles ($Gd_2O_3$:$Er^{3+}$), levitated in a linear electrodynamic trap. These microdroplets served as spherical optical resonators, with their resonance properties influenced by the distribution and interactions of nanoparticles (NPs) acting as nanoprobes. Using a combination of light scattering theories, we examined the evolution of the microdroplet radius, nanoparticle surface organization, and effective refractive index during evaporation. Up-conversion luminescence, driven by 805 nm excitation, enhanced the resolution of these observations. Key findings reveal that the increasing NP density in the surface layer leads to surface saturation, followed by transitions to a gel-like state through successive layer formation and collapse. Scattering and luminescence signals exhibit complex oscillatory behaviors linked to resonance phenomena and nanoparticle migration driven by photophoretic forces. Theoretical modeling closely aligns with experimental trends, confirming the interplay between NP distribution and optical resonance properties.

## 1. Introduction

The study of evaporating microdroplets of colloidal suspensions is crucial in various fields such as material science [1], medical science [2], and climate modelling [3,4]. Evaporation is not only a scientifically important complex process [5,6], but also provides a practical robust method for creating functional nanomaterials used in sensors, optical, and biological devices [7].

The problem of evaporating microdroplets of mixtures is inherently complex due to the possibility of multiple phase changes within the droplet's volume as well as on its surface. Additionally, the evolution of both the surface and volume can be interconnected through volume-surface mixing [8]. As the evaporation progresses and the droplet's radius decreases, the influence of the surface state on the entire droplet increases.

When examining a droplet using light, its spherical shape causes the system to act like an optical resonator (a spherical cavity) with various modes, in particular Whispering Gallery Modes. In this context, the surface of the droplet plays a crucial role in the interactions between the droplet and light [9,10].

In suspension microdroplets, the array of optical phenomena allows for the collection of information about surface states [11], as well as the structures formed by dispersed (nano)particles and their evolution [12].

The thermodynamic states of the surface constitute or at least influence the physicochemical properties of the microdroplet (see e.g. [13]). The microdroplet surface interfaces the environment with the droplet volume, governing the droplet's catalytic properties [14] as well as controlling the mass transport between the surface and the volume.

The surface states influence the drying scenario of the droplet and shape the structure of the dried particle, which is the result of evaporation [15]. The drying scenario has a fundamental impact on aerosol droplets of suspensions, affecting global warming [16], monsoon drying [17], and the contagiousness of pathogens adsorbed in microdroplets produced by respiration, coughing, and sneezing [18].

In previous papers, several properties of the surfaces of evaporating microdroplets of colloidal suspensions were studied, including the mean value of the index of refraction [19], the formation of super plasmons in a dispersion of gold nanoparticles in diethylene glycol (DEG) [20], the formation of surfaces/shells of silica nanoparticles in colloidal suspensions [15], the generation of surface films in SDS/DEG mixtures [21], and Brownian motion of the dispersed nanoparticles [22].

The concept of using various nanoparticles as nanoprobes is already well-grounded – see e.g. [23] for a review. In [24], optical lattices forming in microdroplets containing down-converting luminescent nanoparticles ($Gd_2O_3:Nd^{3+}$) were investigated. In this paper, we examine the evolution of optically measurable properties of single microdroplets of charge-stabilized suspensions containing up-converting luminescent nanoparticles ($Gd_2O_3:Er^{3+}$) as nanoprobes. Since up-conversion is a two-photon process, it enhances the resolution of observations. The study primarily focuses on the droplet radius, density, surface organization of nanoparticles, and the effective refractive index.

A key characteristic of the system is the increasing nanoparticle density in the surface layer, eventually resulting in surface saturation. This progression drives the system toward a gel state through the formation and eventual collapse of successive surface layers.

The observed properties seem to be general for the drying of droplets of colloidal suspensions. These properties can significantly influence the evolution of the physicochemical characteristics of the droplets, the accumulation of substances like drugs or dust, and the evaporation rate of droplets containing dust particles.

## 2. Experimental setup

As presented in Fig. 1, the experiments were conducted in an in-lab built linear electrodynamic quadrupole trap (LEQT; compare [25–27]). The essential trap configuration was the same as described in [24], so only the introduced improvements will be discussed in greater detail. Electrically charged droplets were confined horizontally with the quadrupolar AC field generated by rod electrodes, while vertical confinement was achieved with the DC field of plate electrodes. The field between the plate electrodes, in the configuration we used, is not homogeneous around the centre of the trap, but exhibits a gradient, which enables the translation of microdroplets vertically by varying the applied voltage. The levitating microdroplet was backlit with a bright red LED and its (vertical) position was observed as a shadow with a digital camera (Smartek, GC651MP, equipped with IR filters) through a high-magnification microscope, at 10 fps, with a resolution of ~1.0 µm/pixel. It enabled microdroplet radius measurement (shadowgraphy) as well as setting up a stabilization loop and keeping the droplet at the desired vertical location – on the (horizontal) axis of the light-collecting optics.

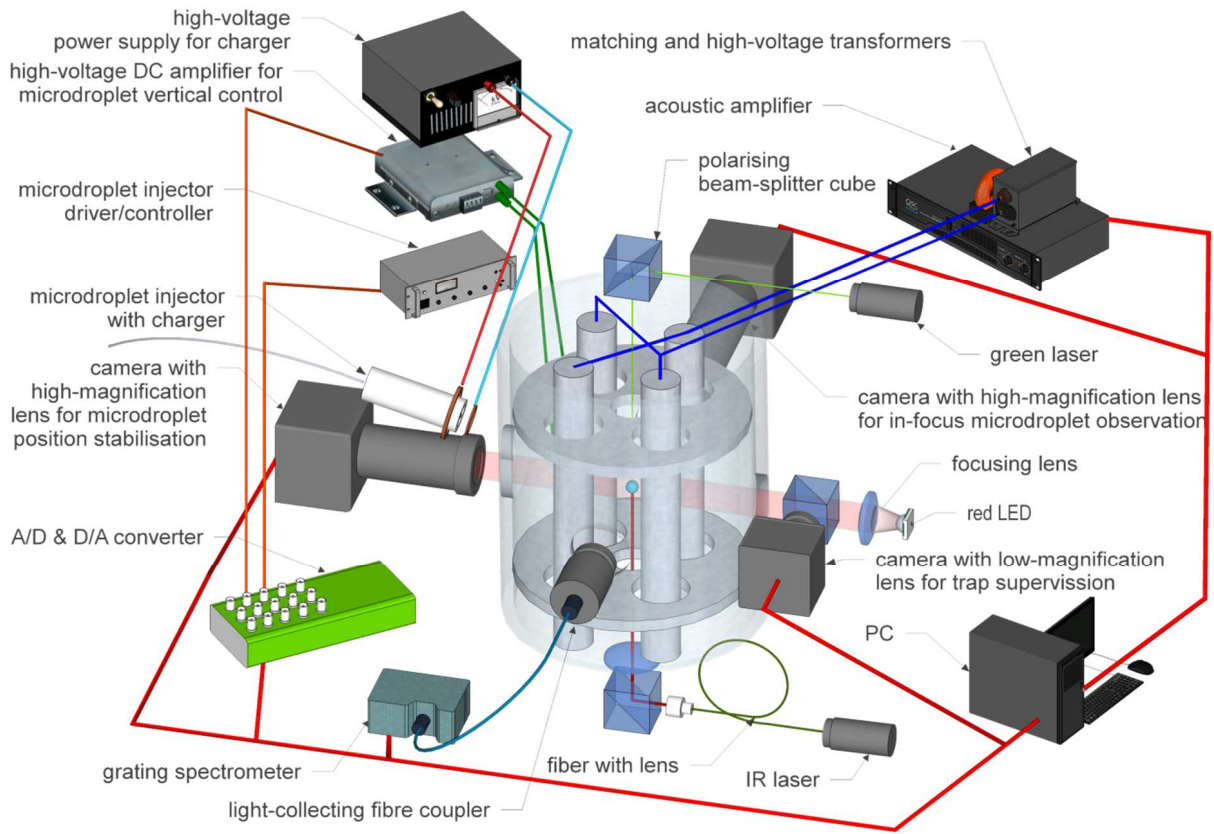

Fig. 1. Schematic diagram of the experimental setup.

$Gd_2O_3$:6%$Er^{3+}$ nanoparticles of 353 nm diameter were suspended in DEG by mixing the dry powder and sonication for several hours until no sediment was present. The mass concentration of NPs was 5 mg/mL, on loading the suspension to the injector. Since the density of $Gd_2O_3$ (~7.1 g/cm$^3$) is very high in comparison to that of DEG (~1.1 g/cm$^3$), the viscosity of DEG (~37 mPa·s) is not high enough to suppress sedimentation. Thus, the initial mass concentration of NPs in the microdroplet, $w_{NP}$, could be slightly different (usually lower). When the suspension was stored, it was kept in a refrigerator in a vertical tube rotator. It was then sonicated for 2–3 hours before the experiments, which were conducted immediately afterward. $Gd_2O_3$:$Er^{3+}$ NPs' synthesis was described in [24] and [28].

Microdroplets were delivered to the trap with an in-lab built droplet-on-demand piezoelectric injector equipped at the nozzle with annular electrodes for droplet charging ([25], compare e.g. [29]). The high voltage used for charging was switched on only for the droplet injection while off directly afterwards, in order not to influence the trapping field during the stabilizing loop operation. In spite of significant effort, due to some sedimentation in the injector, the parameters of microdroplets (radius, charge, initial NS number concentration) varied noticeably, though not significantly.

0In this set of measurements, the microdroplet was illuminated with an IR beam from below, to counteract possible sedimentation of $Gd_2O_3$ nanoparticles in the microdroplet. In order to circularize the beam and, most of all, suppress the spatial fringes produced by multi-junction laser diode, the beam was passed through the multimode 200 μm fibre. The spatial mode structure inherited from the fibre turned out to be much less harmful to the repeatability of the experiment.

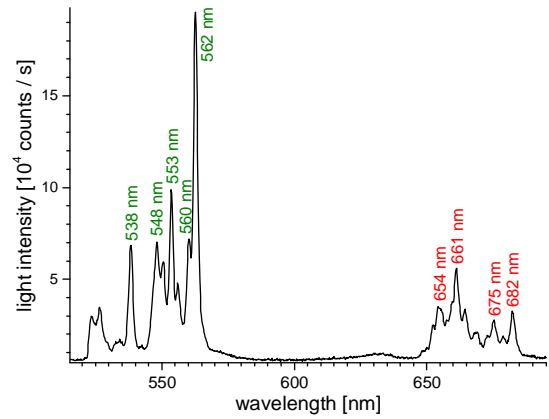

Fig. 2. A fragment of the spectrum – green and red luminescence bands – observed for a microdroplet of 55-μm diameter, containing $Gd_2O_3$:6%$Er^{3+}$ nanoparticles of 508-nm diameter (~40 mg/mL in triethylene glycol). The temporal evolution of the marked lines was followed.

The scattered light and luminescence was collected with a dedicated objective and fed to small grating spectrometer (USB4000, Ocean Optics) via 400 μm fibre. We observed the scattering of laser light at wavelengths of 805 nm (attenuated with two OD6 notch filters) and 515 nm, as well as up-converted luminescence induced by 805 nm radiation, detected at several lines within two bands: red and green – see Fig. 2. Due to low luminescence light intensity, the integration time had to be set to ~1 s. It set the limit to the temporal resolution of the experiments. The experimental data (and spectra in particular) can be accessed at the data repository [30] (in particular sample DI192: 003, 006).

All measurements were performed under ambient conditions, while the lab was air-conditioned with temperature set at 22°C.

## 3. Microdroplet radius

First of all, microdroplet radius found with shadowgraphy shows some meaningful departure from the so called radius-square-law, as can be seen in Fig. 3 (a) and (b). The effects can be observed both for a droplet evolution with a significant, over 3.2:1, radius change (Fig. 3 (a)) as well as for that with only 1.2:1 radius change (Fig. 3 (b)). The radius change dynamics is obviously higher in the first case. A closer investigation of the first case shows that the evaporation rate (radius change rate) d$r$/d$t$ grows (at different pace) until ~5600 s, and then suddenly stabilizes at a significantly lower magnitude. This latest development is easiest to interpret – there forms a compact surface layer of NPs, possibly even a wet aggregate of NPs, which later dries very slowly (compare [15]). The NPs simply cover the surface and block the evaporation.

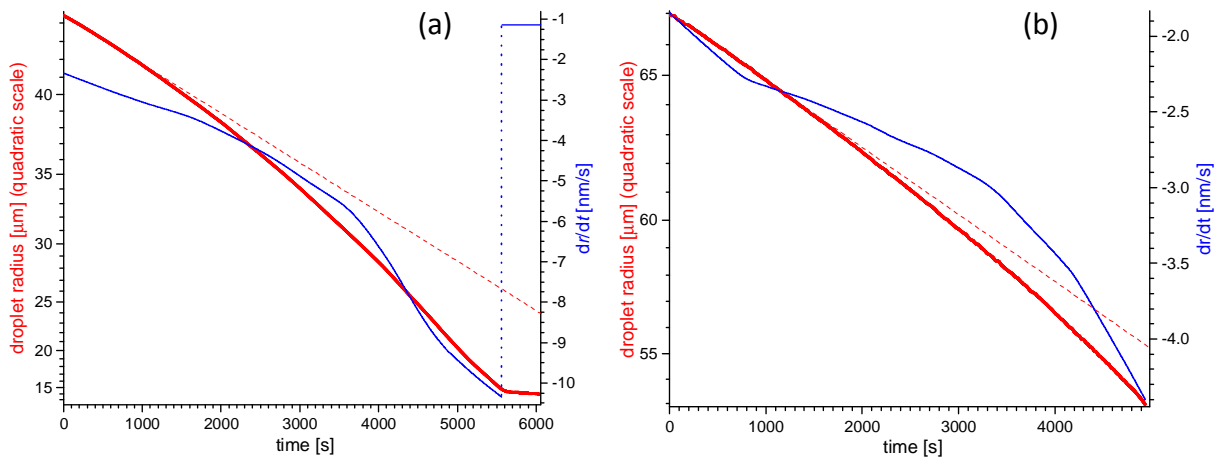

Fig. 3. Evolution of the microdroplet radius (smoothed using Lowess with a 131-point window) and the corresponding radius change rate (smoothed using Lowess with an 899-point window) for two microdroplets of different size ranges. The vertical axes use quadratic scales, so evaporation following the radius-square-law should appear as a straight line, indicated by the red dashed lines.

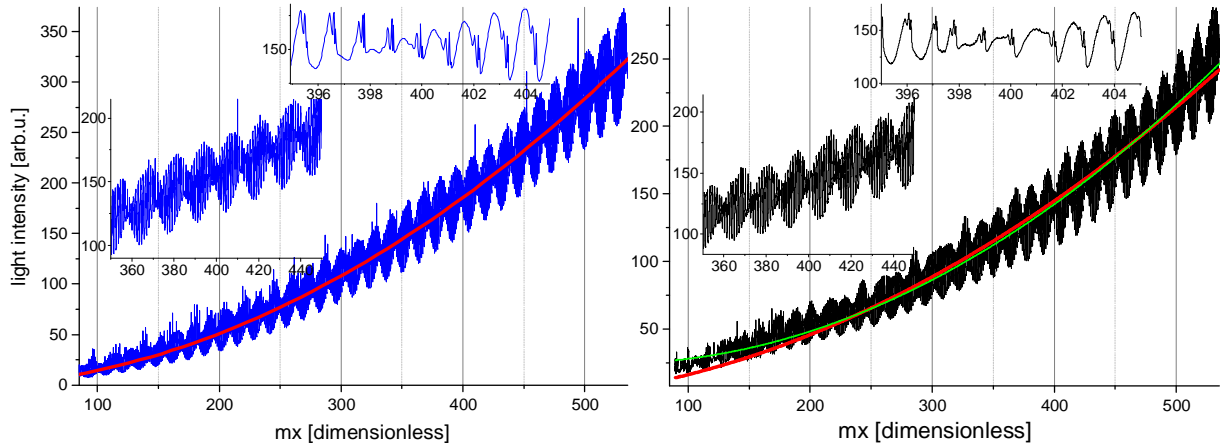

Fig. 4. Scattered light intensity in the far field, integrated over the field-of-view of (90±16° × 0±5°). Left panel – Mie theory calculation, right panel – experiment on pure DEG microdroplet. For both panels: wavelength λ = 458 nm, refractive index of DEG at 22°C and 458 nm $m$ = 1.4524. Radius-square law shown as the red solid line in both panels. The trend of experimental intensity – green solid line – deviates slightly from the radius-square law due to limited purity (99.9 %) of DEG used in the experiment.

Then, it can be also observed that d$r$/d$t$ magnitude exhibits a few "kinks": the first two mark slight accelerations of evaporation rate, while the third – a deceleration. All such kinks seem to correlate with rapid changes of the character of oscillations observed in the optical signals – see Fig. 9. At the same time, it can be observed that as the evaporation rate increases, the frequency of oscillations observed in the optical signals (scattering and luminescence) also increases, as can be generally expected.

Such rapid changes are not predictable with the effective medium approach (compare orange line in Figs. 7 and 9 and violet line in Fig. 11), and require considering rapid changes of the NPs ordering – structural changes of the microdroplet (compare [11]).

Having measured the droplet radius, we calculated light scattering with Mie theory using effective medium approximation and combined with the theory of light scattering by fractal aggregates (see e.g. [31]). In this way, we were able to verify the hypothesis on the NPs distribution evolution. By analysing the evolution of luminescence signals, we studied modifications to the resonance properties of a microdroplet—an inhomogeneous spherical optical resonator—altered or introduced by nanoparticles (NPs) and their distribution.

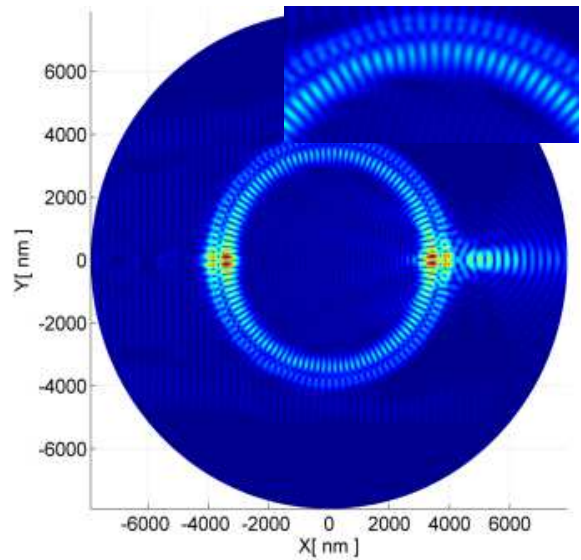

Fig. 5. Light intensity distribution at a narrow resonance predicted by the Mie theory for a homogeneous sphere of $R$ = 3953 nm, $m$ = 1.45 – internal and near field intensity seen in the cross section at the equatorial plane; a magnification of a fragment (X∈[-2000, 2000], Y∈[2000, 4000]) is shown in the inset. Illumination with a plane wave from the left.

## 4. Far-field light scattering by a homogeneous droplet and internal field distribution

The intensity of light scattered by a spherical object, as calculated using Mie theory [32], is proportional to the surface area of the sphere (see e.g.: [33]) and exhibits modulation due to spherical cavity modes. The amplitude of this modulation, which represent a small fraction of the total scattered light intensity, is also proportional to the surface area. Theoretical modeling aligns very accurately with experimental data, such as light scattering observed in the far field during the evaporation of a single droplet of diethylene glycol – see Fig. 4.

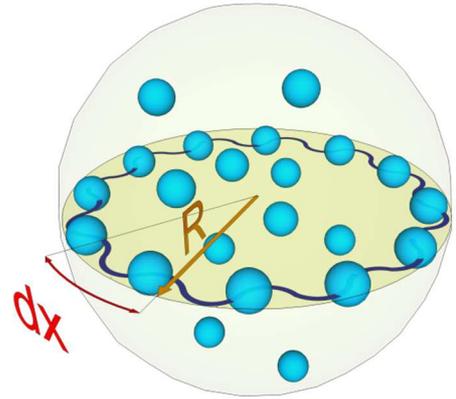

Fig. 6. A conceptualisation of a microdroplet of NPs suspension with a standing light wave interacting with NPs in an equatorial cross-section.

The scattered light intensity maxima observed in the far-field versus the size parameter correspond to resonances of the internal field (see Fig. 7 in [24]; compare e.g. [34]). Internal-field simulation using Mie theory shows that the light intensity is concentrated near the interface (droplet surface) and is modulated spatially with the distance around the circumference – see Fig. 5 (compare e.g. [35,36]).

## 5. Intra-cavity field distribution and NPs distribution

The radiation intensity inside a homogeneous microsphere with a linear refractive index can be rigorously determined using Mie theory (see e.g.: [32,37]). However, for a spherical microdroplet containing a suspension of luminescent (nano)particles that absorb and scatter light, this provides only a rough approximation. Some phenomena can be grasped with effective medium approach – as we show below, while an exact solution to the problem remains out of reach for now. However, to grasp/visualize the resonance properties of a spherical cavity, it turns out sufficient to consider a 2D model of a circular resonator [38] – see Fig. 6 (compare Fig. 5). Thus, the intensity of radiation (light) at the circumference of a spherical cavity (homogeneous droplet) in the equatorial plane can be simplistically described as an intra-cavity standing wave:

$$I(R,x) \sim 1 + \cos(\Omega_R R)\cos(\Omega_x x), \qquad (1)$$

where $R$ is the decreasing in time droplet radius and $x$ is the position along the circumference. It gives an intuitive picture of a complex distribution. The frequencies $\Omega_R$ and $\Omega_x$ can be determined by considering the interval/spacing between the consecutive intensity maxima/minima versus $R$ and $x$: $dR = \lambda/2\pi m$ and $dx = \lambda/m$, where $\lambda$ is the vacuum wavelength of radiation and $m$ is the refractive index of the droplet. The obvious condition: $\Omega_x dx = \Omega_R dR = 2\pi$ leads to $\Omega_R = 4\pi^2 \lambda/m = 2\pi\Omega_x$. Any phase relations are of no importance in this description.

As mentioned before, NPs dispersed in a microdroplet interact with its internal field (absorb, scatter), and thus modify it. In modelling of scattering/absorption/luminescence on a microdroplet as a whole, the distribution of NPs – particularly near the surface, where the field is strongest at resonance – must be taken into account in some way, and therefore must also be modelled/predicted beforehand.

Since the NPs dispersion is usually charge-stabilised their distribution naturally evolves from completely random to somehow organised (e.g. hexagonal in layers), in particular for the higher NPs number density. We assume that the NPs distribution in the microdroplet surface layer (most important to interactions with radiation) can be approximated as Gaussian, characterized by the mean distance $\mu$ and standard deviation $\sigma$ (distribution width):

$$N(x) = \frac{N_{\text{surf}}}{\sigma\sqrt{2\pi}} \exp\left[-\left(\frac{x-\mu}{\sqrt{2}\sigma}\right)^2\right], \qquad (2)$$

where $N_{\text{surf}}$ is the number of NPs in the surface layer and $\sigma$ is obviously much smaller than the droplet circumference $2\pi R$. When NPs arrange themselves in a regular pattern – transitioning from a "surface gas" to a "surface solid" phase, $\sigma$ becomes smaller.

As mentioned above, just after the droplet formation, at the beginning of the evolution, the number density $n$ of NPs in the droplet can be considered uniform – equal at the surface and in the droplet volume $n_{\text{surf}} = n$. Due to the evaporation of the dispersion medium, the droplet surface is moving radially and capturing particles from the evaporated volume. Additionally, the existence of surface adsorption and charge of NPs helps to move them to the interface until all radial forces acting on NPs are equilibrated [39] and most of the nanoparticles reside at or close to the interface. The details of the evolution depend on the relation between the mentioned processes. In this consideration, $N_{\text{surf}}$ was simplistically modelled as resulting only from the droplet evaporation – interface inward movement:

$$N_{\text{surf}}(R) \cong \frac{w_{\text{NP}}(R_0^3 - R^3)}{\rho_{\text{NP}} r^3}, \qquad (3)$$

where $R_0$ is the initial droplet radius, $w_{\text{NP}}$ is the initial mass concentration of NPs, $r$ is the NP radius and $\rho_{\text{NP}}$ is the density of NPs' material. This formula should reasonably describe the number of NPs at the surface until the surface saturation, while after the surface collapse(s), it should give the total number of NPs in the near-interface layers. We shall verify this assumption in Section 6.1 – the black dotted line in Fig. 7 represents $N_{\text{surf}}$ corresponding to the radius evolution shown in Fig. 3 (a).

On the other hand, light-absorbing NPs are expected to migrate to the minima of absorbed light intensity due to photophoretic forces, rather than to the maxima via gradient forces, thereby reflecting the temporal evolution of these minima. In the discussed experiments, the 805-nm radiation absorbed in $Gd_2O_3$:$Er^{3+}$ NPs drives a "standing wave" of NPs density, which feedbacks with 805-nm internal field distribution and heavily influences scattering and luminescence at all wavelengths. Furthermore, this "standing wave" of NPs density modifies resonance properties to the microdroplet – introduces additional resonances and modifies the existing ones (compare e.g. [40,41]). This phenomenon can be clearly observed in Fig. 10 as a transition in the modulation of scattered light intensities at different wavelengths. Initially, the modulation corresponds to spherical cavity modes specific to each wavelength, characterized versus $R$ by a period of $1/\Omega_R = m/4\pi^2\lambda$. Over time, this transitions to a unified period corresponding to the 805 nm line. We discuss it below in detail.

## 6. Scattering of light on a droplet of suspension

Having in mind the considerations of the previous section, several hypotheses were tested using Mie theory-based codes, and we found that the increase of effective refractive index of the microdroplet can explain the observed trend (long-time average) of the scattered light intensity (see the cyan line in Fig. 7), but not the oscillatory properties of the signal. A separate term describing the scattering on NPs is necessary (compare Figs. 7 and 11). Using the analysis for the corresponding characteristic size relationship (spatial coherence criterion), presented e.g. in [31], we've arrived at conclusion that the scattering on NPs is proportional to number of NPs at the surface and must also depend on their distribution $N(x)$. In order to fully explain the oscillatory properties of the signal, the second term must be modulated by the internal field of the droplet (we shall discuss the modulation of scattering and luminescence signals in detail in section 6.3). Further on the correlation between $N(x)$ and $I(x)$ should be accounted for. Thus, the intensity of light of wavelength $\lambda$, scattered on a microdroplet of radius $R$ could be postulated as:

$$I_{\text{sca}}(\lambda, R) \cong I_{\text{eff}}\big(\lambda, R, m_{\text{eff}}(R)\big) + \alpha \int_0^{2\pi R} N(R, x) I(R, x) \mathrm{d}x ,  \qquad (4)$$

where the proportionality factor $\alpha$ depends, among others, on the microdroplet and NP size parameter, and the composite microdroplet mass and surface fractal dimensions. $I_{\text{eff}}$ is the intensity of light scattered by an effective homogenous sphere with an effective refractive index $m_{\text{eff}}$, which follows a simple mixing rule for refractive indices of $Gd_2O_3$ and DEG:

$$m_{\text{eff}}(R) = c_{\text{Gd2O3}} \cdot m_{\text{Gd2O3}} + (1 - c_{\text{Gd2O3}}) \cdot m_{\text{DEG}} , \qquad (5)$$

where, in turn, $c_{\text{Gd2O3}}$ is the volume fraction of NPs residing in the near-surface layer of the droplet:

$$c_{\text{Gd2O3}} = r^2 N_{\text{surf}} / 4R^2 . \qquad (6)$$

The required change in refractive index actually occurs in the near-surface layer of the droplet due to the accumulation of NPs there, which is the result of liquid evaporation. The value of the refractive index of the droplet interior was found of less importance, if not entirely negligible.

### 6.1 NPs distribution evolution and its interaction with light

Now, it must be noticed that the second term in the Eqn. (4) does not hold well when $N(R,x)$ is itself driven by the strong internal field at 805 nm line. In that case, we postulate a semi-empirical formula:

$$I_{\text{sca}}(\lambda, R) \cong I_{\text{eff}}\big(\lambda, R, m_{\text{eff}}(R)\big) + \alpha N_{\text{surf}} I_{\text{eff}}\big(805, R + \beta N_{\text{surf}}^2, m_{\text{eff}}(R)\big) , \qquad (7)$$

which we will discuss in detail in the following sections. Again, the proportionality factors $\alpha$, $\beta$ depend, among others, on the microdroplet and NP size parameter, and the composite microdroplet mass and surface fractal dimensions. The factors, we found for scattering at different wavelengths, were consistent.

First of all, at this point it, is possible to verify the hypothesis on the $N_{\mathrm{surf}}(R)$ evolution (Eqn. (3)). If we neglect the influence of the refractive index change/evolution and assume the long-time trend of $I_{\mathrm{eff}} \propto R^2$, and we observe that the dynamics of Eqn. (7) is dominated by $N_{\mathrm{surf}}(t)$, we can write:

$$I_{\mathrm{sca}}(R) \approx R^2 + \alpha N_{\mathrm{surf}}. \quad (8)$$

Then, from the long-term temporal variability of the flux density $i_{\mathrm{sca}} = I_{\mathrm{sca}}/R^2$, we should be able to grasp the main trends of the evolution of the surface NPs' number density: $n_{\mathrm{surf}} \propto i_{\mathrm{sca}}$. As can be seen in Fig. 8, the $n_{\mathrm{surf}}$ evolution obtained with Eqn. (3) is in a very good agreement with $i_{\mathrm{sca}}$ obtained from the experiment.

It must be kept in mind that undulations of the optical signals in Fig. 8 arise rather due to specific interactions of light with NP structure/distribution (surface-thermodynamics states) [19,42], than to significant short-time variations of $n_{\mathrm{surf}}$. Perhaps with the exception of the collapse(s) of the NPs surface layer observed in Fig. 8 as deep oscillations near the end of the evolution. It can be further noticed that different wavelength – different field modes – interact differently with the microdroplet with evolving NPs distribution. It can be noticed, for instance, that effectiveness of scattering of the non-coherent unpolarized (red) light from an LED $i_s(632)$ exhibits fairly little interaction with NPs until ~3600 s, when the number of NPs at the droplet surface starts to grow. The $i_s(805)$ of IR coherent light, which is absorbed by NPs (which manifests in smaller modulation depth), seems to detects NPs somewhat deeper below the droplet surface and thus its increase starts earlier – practically from the beginning. Finally, $i_s(515)$ – non-absorbed, coherent light – additionally shows some interaction with the transient NP structures forming in the microdroplet. It can be seen both as broad maximum (maxima?) as well as high modulation depth.

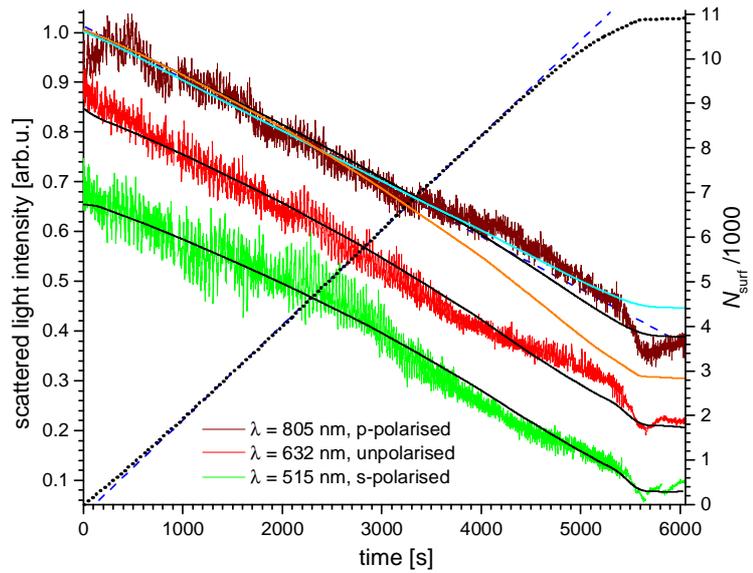

Fig. 7. Scattered light intensity at three different wavelengths and polarisations. This incident light also differs in character: only 805-nm light is absorbed in NPs; 632-nm light is non-coherent. The scale for each trace is independent. The solid black lines are long-time averages predicted with Eqn. 7. The orange and cyan lines correspond to $R(t)^2$ and the first term of Eqn. 7, respectively. The dotted black line shows the predicted/modelled evolution of $N_s$. The dashed blue lines are eye-guiding straight lines.

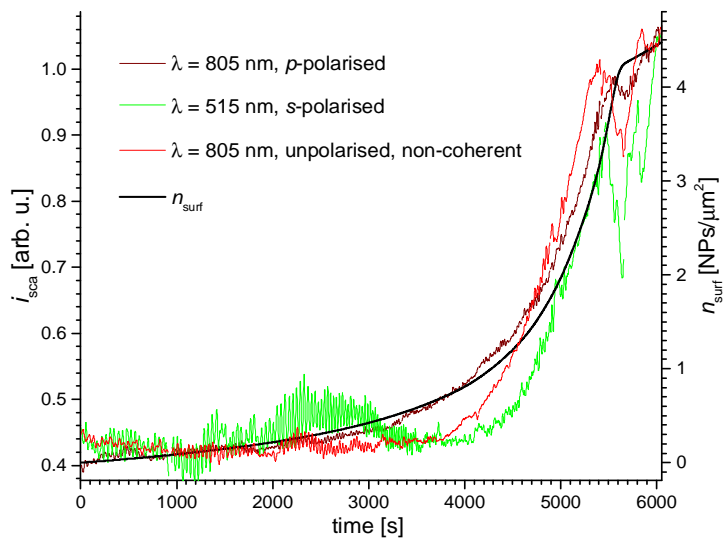

Fig. 8. Scattered light flux density at three wavelengths and the modelled surface number density of NPs. The $i_{\mathrm{sca}}$ scales at different wavelengths are independent.

## 6.2 Scattering long-time trend

It could be expected that the intensity of scattered light – its long-time trend – decreases in a similar manner as the square of the radius (see Fig. 3 (a)) – orange solid line in Fig. 7. However, this is not the case, and the observed decrease resembles linear – see blue dashed line in this figure. This is in line with what has been observed for the oscillatory part of the signal – see below – it indicates that there is a process, which slows the evaporation down. As it has been established, it is the presence of suspended phase – NPs – that governs the phenomenon. NPs located in regions of a strong field (e.g., close to the surface) are expected to contribute the most to the total scattering. It turns out, that the averaged Eqn. (7) predicts the observed trends quite well, as can be seen in Fig. 7.

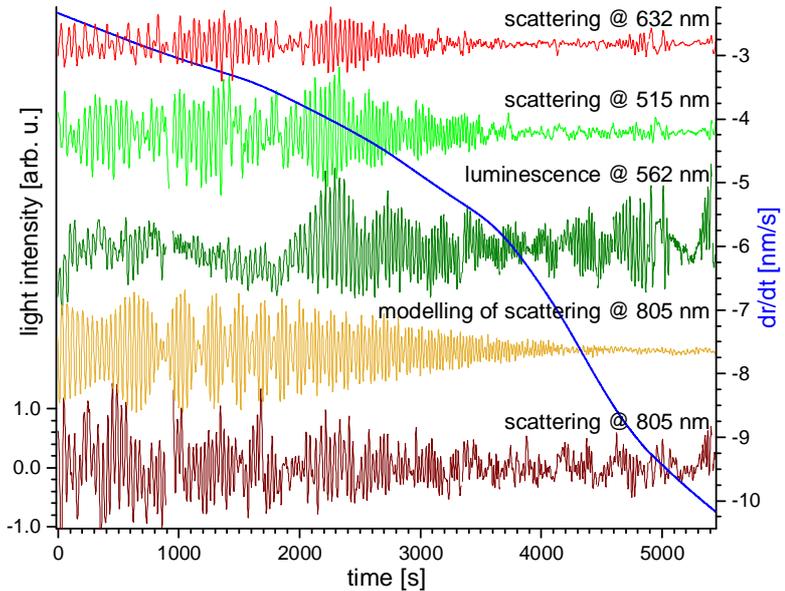

fig. 9. Oscillatory part of optical signals (see main text) shown with independent scales. Microdroplet radius change rate is shown in blue solid line.

## 6.3 Oscillatory part of scattering and luminescence signals

However, the factors discussed so far are not by themselves sufficient to explain the properties of the oscillatory part of the signals, among others: (i) There are striking transitions from de-phased signals oscillating at individual eigenfrequencies, to synchronised oscillations with a common frequency – see e.g. transition at ~2050 s in Fig. 10. This synchronicity is lost at ~3500 s. Synchronisation repeats at ~4600 s and lasts only until ~4900 s. They seem to be associated with formation and decay of some regular NP structures at the droplet surface (compare [43]). (ii) The mean frequency of all signals decreases much more slowly than predicted only by the first term of equation (7) (effective homogeneous droplet) – compare the purple line in Fig. (peak positions).

Again, several hypotheses were tested and, as mentioned in section 3, the migration/ordering of NPs caused by intra-cavity 805 nm radiation was proposed as the common factor influencing both scattering and luminescence on all spectral lines. Thus, it seemed justified to introduce $N_{surf}$ modulation with $I_{eff}(\lambda = 805$ nm$)$, as resonances observed in the internal and the scattered fields are very simi-lar, as discussed earlier. Furthermore, since it can be well expected that the migration of NPs introduces some lag in respect to the driving radiation intensity due to viscosity, some retardation had to be introduced into the second term. We found out that it is proportional to $N_{surf}^2$, which seems to indicate the dependence on the NPs surface lattice stiffness (NPs mutual interactions). The interference of the two terms generates beating, which can also be identified in the experimental data, as can be seen in Fig. 9. Since the specific resonance features of the system for each wavelength are neglected, the reproduced beating shows only a general property.

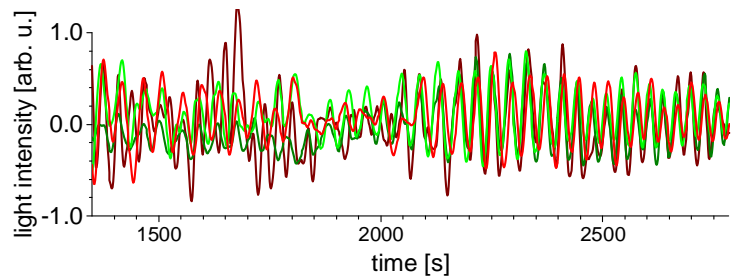

Fig. 10. Transition from de-phased signals oscillating at individual eigenfrequency, to synchronised oscillations with a common frequency, at ~2050 s.

The oscillatory part of the signals can be conveniently studied when the slow trend is subtracted from the raw signal, and – in case of experimental data – some smoothing is performed, which filters out the random fast component of the signal – see Fig. 9. This fast-changing component could not be further resolved in the reported experiment (as mentioned in section 2) due to sensitivity limitations of the spectrometer. Thus, due to possible aliasing, it could not be decided whether deformations and movements of the droplet caused by the trapping field are reflected in this random signal component or it was just caused by random inhomogeneity of the microdroplet. In the next step, the extrema (e.g. maxima) in this processed signal can be identified and/or frequency analysis (short time Fourier, wavelet) can be performed. We've decided that analysis of maxima positions presents a convincing visualisation of the observed phenomena – see Fig. 11.

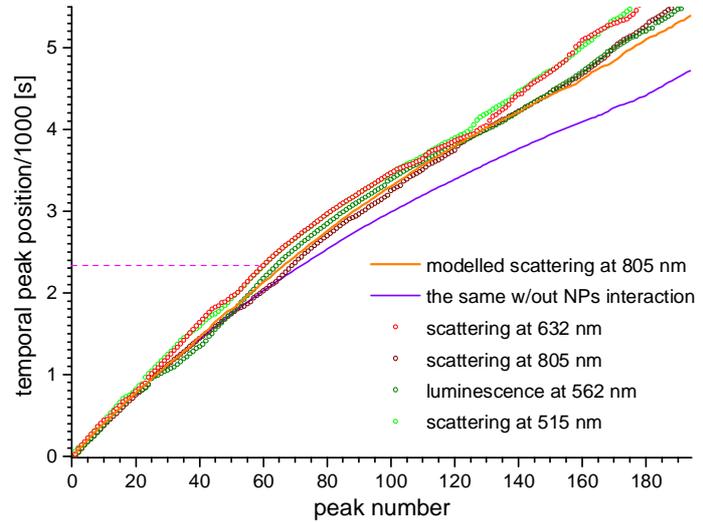

Fig. 11. Temporal positions of subsequent peaks seen in Fig. 9. Color-coding corresponding. The violet solid line corresponds to prediction with the first term of Eqn. 7 only. Horizontal dashed line marks the approximate position peak frequency change.

It can be observed in Fig. 11 that frequency of oscillations of 805 nm scattering (and so essentially also of internal field) and 562 nm luminescence is nearly identical (which is also replicated for other luminescence signals). This oscillations frequency evolution has been quite well reproduced with Eqn. (7) – orange solid line in Fig. 11. As mentioned above, the violet line in Fig. 11 corresponds to the first term of this equation only. The oscillations of 632 nm scattering of unpolarized non-coherent light from an LED and scattering of 515 nm coherent p-polarized 515 nm laser light are again nearly similar and both are quite similar to 805 nm filed oscillations. It should be kept in mind that 805 nm radiation is absorbed in NPs, while other wavelengths, considered in the presented experiments, are not. It can be also observed in Fig. 9 that each of the presented signals exhibit somewhat different modulation, which signifies different interactions with inhomogeneous spherical resonator – NPs distribution and the microdroplet as a whole. Its evolving structure of morphology dependent resonances (MDRs) is "scanned" with light of different wavelengths.

Here, it is worth noting as a side remark that the light at 805 nm (the excitation radiation) interacts with light at other wavelengths by modifying the medium (i.e., the distribution of nanoparticles). Under such conditions, the microdroplet can be perceived as optically non-linear, enabling light-light interaction.

## 7. Intensity of luminescence long-time trend

The intensity of luminescence (up-conversion) appears to be simpler to analyse compared to scattering, as it primarily depends on the intensity of the excitation light (at 805 nm) and the correlation with the distribution of particles:

$$I_{lumi} \sim \int_0^{2\pi R} N(x) I^2(\lambda = 805 \text{ nm}, x) \mathrm{d}x . \tag{9}$$

It can be divided into two components: one that is proportional to the number of illuminated nanoparticles and an oscillatory component that reflects how well the excitation radiation spherical cavity (droplet) mode aligns with the mean distance between the NPs. Since luminescence is driven by 805 nm intra-cavity radiation, it also inherits the oscillatory properties predicted with equation (7). The long-term signal variation requires additional consideration. It should be also kept in mind that for the up-conversion driven luminescence, $I_{lumi} \propto I_0^2$, where $I_0$ is the incident light intensity.

Under the reasonable assumption that the quantum efficiency of $Er^{3+}$ luminescence in $Gd_2O_3$ lattice can be considered constant, the luminescence signal must depend directly on the number of excited $Er^{3+}$ ions. Since the concentration of Er in $Gd_2O_3$ is constant for a given sample, the luminescence signal $I_{lumi}$ must then depend linearly on the (total) number of NPs $N_{tot}$. Since here we consider a long-time trend, it is enough to consider that the energy of the 805 nm radiation that can be passed to $Er^{3+}$ is just proportional to the geometric cross-section of the microdroplet. While $N_{tot}$ is obviously constant, the cross-section of the microdroplet evolves in time as $R^2(t)$.

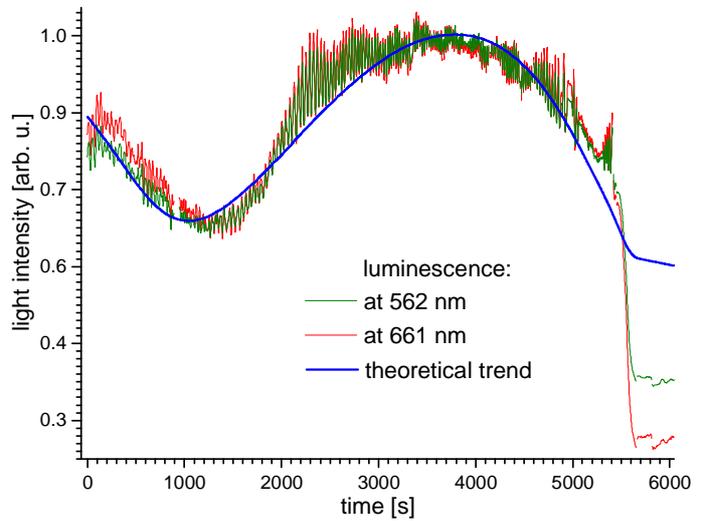

Fig. 12. Luminescence at two spectra lines resulting from different up-conversion mechanisms, together with the prediction with Eqn. (10).

However, these considerations do not explain the broad maximum seen in the luminescence signals and exactly overlapping on all luminescence spectral lines. This maximum seems to signify some resonant feature of the system. Furthermore, the resonance must be attributed to a single wavelength, since there is no temporal shift between signals on different wavelength. It can be well expected that again it must be attributed to 805 nm radiation (excitation). In line with our previous works [20,42], we expect that the distributed feedback/additional resonance introduced by the specific NPs distribution strongly enhances two-photon excitation (up-conversion). This effect cannot be identified in scattering, which is dominated by scattering on the droplet as a whole, while also it is a single-photon process.

Thus, it makes sense to compare the mean distance between NPs with the wavelength and/or distance between field maxima/minima on the resonator circumference. Here, comparing the distribution of NPs described by Eqn. (2) with the excitation radiation wavelength in the (effective) medium λ yielded the expected results. Thus, the intensity of luminescence could be approximated as

$$I_{lumi} = aR^2 N_{tot} + b \frac{N_{surf}}{\sigma\sqrt{2\pi}} \exp\left(-\left(\frac{x(R)-\lambda}{\sqrt{2}\sigma}\right)^2\right), \tag{10}$$

where, *a* and *b* are constants. We confirmed with the fit (see Fig. 12) that $N_{surf}$ must be, in a way, accounted for twice: the first term must contain $N_{tot}$, rather than the number of NPs only below the surface: $N_{tot} - N_{surf}$, and the second term accounts for an additional resonant feature.

It can be observed that the peak value of the resonance maximum in the luminescence signal depends on the width of the particle distribution σ. Similar effect manifests for interaction of light field with regularly structured NPs distribution. It can be observed at ~2000 s and ~4600 s in Fig. 9 as an increase of luminescence modulation amplitude. Therefore the amplitude of luminescence intensity modulation could serve as a measure of σ.

Notably, in the language of surface thermodynamics, the width of the particle distribution σ is related to the entropy of the particle layer at the surface (see e.g. [44]):

$$H = \frac{1}{2}\log(2\pi\sigma^2) + \frac{1}{2}. \tag{11}$$

This connection should allow us to quantify the disorder or randomness of the particle arrangement on the surface by measuring the amplitude of luminescence signal modulation.

## 8. Conclusion

This study investigates the dynamic optical properties of single evaporating microdroplets containing up-converting luminescent nanoparticles ($Gd_2O_3$:$Er^{3+}$). The microdroplets, acting as spherical optical resonators, enable the exploration of the intricate interplay between light-matter interactions, resonance phenomena, and the evolving nanoparticle distribution.

The resonant modes of the droplet, particularly whispering gallery modes, form standing wave patterns within the droplet that dynamically interact with the nanoparticle arrangements, modulating the scattered and luminescent light intensities. These modes can be conceptualized as "optical atom" states, where the light field within the resonator forms periodic, quantized distributions akin to electronic orbits in an atom [9,45]. Deviations from perfect symmetry, caused by nanoparticle inclusions or surface roughness, break the degeneracies of these modes, further shaping the optical behavior (compare [46]).

The evaporation process concentrates nanoparticles at the droplet surface, leading to increasing surface density, saturation, collapse(s), and eventual transitions into a gel-like state. This evolution alters the resonance properties, introducing new optical modes and modifying existing ones. Photophoretic forces, exerted by the strong internal field (excitation at 805 nm), play a significant role in driving nanoparticles toward regions of light intensity minima within the droplet. This, in turn, influences resonance dynamics and enhances the structural complexity of these "optical atom" states.

Theoretical modeling aligns well with experimental results, validating the hypothesis that nanoparticle distribution significantly impacts the droplet's optical characteristics. The interplay between the evolving physical properties of the droplet and its resonant optical states provides a deeper understanding of light-matter interactions in such systems.

These findings underscore the intricate relationship between optical resonance phenomena and the evolving physical properties of microdroplets. The study's outcomes have implications for understanding aerosol behavior, nanoparticle self-assembly, and evaporation dynamics in colloidal systems. Further exploration of light-matter interactions in multi-component droplets could provide deeper insights into the physicochemical processes underlying these phenomena.

# Declaration of generative AI and AI-assisted technologies in the writing process

During the preparation of this work the authors used ChatGPT 4o in order to interactively compose Abstract, Conclusion and Highlights, and to improve the style of several paragraphs in other sections. After using this tool/service, the authors reviewed and edited the content as needed and take full responsibility for the content of the published article.

# Acknowledgments

This research was funded in whole or in part by National Science Centre, Poland, grant 2021/41/B/ST3/00069. For the purpose of Open Access, the author has applied a CC-BY public copyright licence to any Author Accepted Manuscript (AAM) version arising from this submission.

# References

[1] V.M. Shalaev, ed., Optical Properties of Nanostructured Random Media, Springer Berlin Heidelberg, Berlin, Heidelberg, 2002. https://doi.org/10.1007/3-540-44948-5.

[2] G. Iannacchione, A. Pal, Bio-colloidal Drying Droplets: Current Trends and Future Perspectives on Image Processing Applications, Acad. Lett. (2021). https://doi.org/10.20935/al2661.

[3] A. Arakawa, J.H. Jung, C.M. Wu, Toward unification of the multiscale modeling of the atmosphere, Atmos. Chem. Phys. 11 (2011) 3731–3742. https://doi.org/10.5194/acp-11-3731-2011.

[4] W.K. Tao, M.W. Moncrieff, Multiscale cloud system modeling, Rev. Geophys. 47 (2009) 1–41. https://doi.org/10.1029/2008RG000276.

[5] D. Zang, S. Tarafdar, Y.Y. Tarasevich, M. Dutta Choudhury, T. Dutta, Evaporation of a Droplet: From physics to applications, Phys. Rep. 804 (2019) 1–56. https://doi.org/10.1016/j.physrep.2019.01.008.

[6] M.E. Diveky, S. Roy, J.W. Cremer, G. David, R. Signorell, Assessing relative humidity dependent photoacoustics to retrieve mass accommodation coefficients of single optically trapped aerosol particles, Phys. Chem. Chem. Phys. 21 (2019) 4721–4731. https://doi.org/10.1039/C8CP06980H.

[7] B.A. Parviz, D. Ryan, G.M. Whitesides, Using Self-Assembly for the Fabrication of Nano-Scale Electronic and Photonic Devices, IEEE Trans. Adv. Packag. 26 (2003) 233–241. https://doi.org/10.1109/TADVP.2003.817971.

[8] M.A. Bruning, C.D. Ohl, A. Marin, Soft cavitation in colloidal droplets, Soft Matter 17 (2021) 1861–1872. https://doi.org/10.1039/d0sm02002h.

[9] R. Symes, R.M. Sayer, J.P. Reid, Cavity enhanced droplet spectroscopy: Principles, perspectives and prospects, Phys. Chem. Chem. Phys. 6 (2004) 474. https://doi.org/10.1039/b313370b.

[10] L. Cai, J. Pan, Y. Zhao, J. Wang, S. Xiao, Whispering Gallery Mode Optical Microresonators: Structures and Sensing Applications, Phys. Status Solidi Appl. Mater. Sci. 217 (2020) 1–18. https://doi.org/10.1002/pssa.201900825.

[11] D. Jakubczyk, M. Kolwas, G. Derkachov, K. Kolwas, Surface States of Microdroplet of

Suspension, J. Phys. Chem. C 113 (2009) 10598–10602. https://doi.org/10.1021/jp9007812.

[12] M. Kolwas, K. Kolwas, G. Derkachov, D. Jakubczyk, Surface diagnostics of evaporating droplets of nanosphere suspension: Fano interference and surface pressure, Phys. Chem. Chem. Phys. 17 (2015) 6881–6888. https://doi.org/10.1039/c5cp00013k.

[13] A.W. Adamson, T.A. Adamson, A.P. Gast, Physical Chemistry of Surfaces, Wiley, 1997. https://books.google.pl/books?id=rK4PAQAAMAAJ.

[14] M. Ryckelynck, S. Baudrey, C. Rick, A. Marin, F. Coldren, E. Westhof, A.D. Griffiths, Using droplet-based microfluidics to improve the catalytic properties of RNA under multiple-turnover conditions, Rna 21 (2015) 458–469. https://doi.org/10.1261/rna.048033.114.

[15] M. Woźniak, G. Derkachov, K. Kolwas, J. Archer, T. Wojciechowski, D. Jakubczyk, M. Kolwas, Formation of Highly Ordered Spherical Aggregates from Drying Microdroplets of Colloidal Suspension, Langmuir 31 (2015) 7860–7868. https://doi.org/10.1021/acs.langmuir.5b01621.

[16] U. Lohmann, F. Friebel, Z.A. Kanji, F. Mahrt, A.A. Mensah, D. Neubauer, Future warming exacerbated by aged-soot effect on cloud formation, Nat. Geosci. 13 (2020) 674–680. https://doi.org/10.1038/s41561-020-0631-0.

[17] Z. Li, W.K.-M. Lau, V. Ramanathan, G. Wu, Y. Ding, M.G. Manoj, J. Liu, Y. Qian, J. Li, T. Zhou, J. Fan, D. Rosenfeld, Y. Ming, Y. Wang, J. Huang, B. Wang, X. Xu, S. -S. Lee, M. Cribb, F. Zhang, X. Yang, C. Zhao, T. Takemura, K. Wang, X. Xia, Y. Yin, H. Zhang, J. Guo, P.M. Zhai, N. Sugimoto, S.S. Babu, G.P. Brasseur, Aerosol and monsoon climate interactions over Asia, Rev. Geophys. 54 (2016) 866–929. https://doi.org/10.1002/2015RG000500.

[18] M. Rezaei, R.R. Netz, Airborne virus transmission via respiratory droplets: Effects of droplet evaporation and sedimentation, Curr. Opin. Colloid Interface Sci. 55 (2021) 101471. https://doi.org/10.1016/j.cocis.2021.101471.

[19] D. Jakubczyk, G. Derkachov, W. Bazhan, E. Łusakowska, K. Kolwas, M. Kolwas, Study of microscopic properties of water fullerene suspensions by means of resonant light scattering analysis, J. Phys. D. Appl. Phys. 37 (2004) 2918–2924. https://doi.org/10.1088/0022-3727/37/20/021.

[20] M. Kolwas, K. Kolwas, D. Jakubczyk, G. Derkachov, Collective scattering of light on gold nanospheres dispersed in diethylene glycol microdroplet, Acta Phys. Pol. A 131 (2017). https://doi.org/10.12693/APhysPolA.131.288.

[21] M. Kolwas, D. Jakubczyk, J. Archer, T. Do Duc, Evolution of mass, surface layer composition and light scattering of evaporating, single microdropletes of SDS/DEG suspension. Shrinking droplet surface as the micelles generator, J. Quant. Spectrosc. Radiat. Transf. 258 (2021) 107396. https://doi.org/10.1016/j.jqsrt.2020.107396.

[22] G. Derkachov, D. Jakubczyk, K. Kolwas, K. Piekarski, Y. Shopa, M. Woźniak, Dynamic light scattering investigation of single levitated micrometre-sized droplets containing spherical nanoparticles, Measurement 158 (2020) 107681. https://doi.org/10.1016/j.measurement.2020.107681.

[23] S. Sreejith, J. Ajayan, J.M. Radhika, N. V. Uma Reddy, M. Manikandan, Recent advances in nano biosensors: An overview, Meas. J. Int. Meas. Confed. 236 (2024). https://doi.org/10.1016/j.measurement.2024.115073.

[24] Y. Shopa, M. Kolwas, I. Kamińska, G. Derkachov, K. Nyandey, T. Jakubczyk, T. Wojciechowski, A. Derkachova, D. Jakubczyk, Luminescent nanoparticles in a shrinking spherical cavity – probing the evaporating microdroplets of colloidal suspension – optical

lattices and structural transitions, J. Quant. Spectrosc. Radiat. Transf. 296 (2023) 108439. https://doi.org/10.1016/j.jqsrt.2022.108439.

[25] M. Woźniak, J. Archer, T. Wojciechowski, G. Derkachov, T. Jakubczyk, K. Kolwas, M. Kolwas, D. Jakubczyk, Application of a linear electrodynamic quadrupole trap for production of nanoparticle aggregates from drying microdroplets of colloidal suspension, J. Instrum. 14 (2019) P12007--P12007. https://doi.org/10.1088/1748-0221/14/12/p12007.

[26] E.J. Davis, A History of Single Aerosol Particle Levitation, Aerosol Sci. Technol. 26 (1997) 212–254. https://doi.org/10.1080/02786829708965426.

[27] M.B. Hart, V. Sivaprakasam, J.D. Eversole, L.J. Johnson, J. Czege, Optical measurements from single levitated particles using a linear electrodynamic quadrupole trap, Appl. Opt. 54 (2015) F174. https://doi.org/10.1364/ao.54.00f174.

[28] I. Kamińska, D. Elbaum, B. Sikora, P. Kowalik, J. Mikulski, Z. Felcyn, P. Samol, T. Wojciechowski, R. Minikayev, W. Paszkowicz, W. Zaleszczyk, M. Szewczyk, A. Konopka, G. Gruzeł, M. Pawlyta, M. Donten, K. Ciszak, K. Zajdel, M. Frontczak-Baniewicz, P. Stępień, M. Łapiński, G. Wilczyński, K. Fronc, Single-step synthesis of Er3+ and Yb3+ ions doped molybdate/Gd2 O3 core–shell nanoparticles for biomedical imaging, Nanotechnology 29 (2018) 025702. https://doi.org/10.1088/1361-6528/aa9974.

[29] N. Riefler, T. Wriedt, U. Fritsching, Flexible Piezoelectric Drop-On-Demand Droplet Generation, in: Proc. ILASS–Europe 2017. 28th Conf. Liq. At. Spray Syst., Universitat Politècnica València, Valencia, 2017. https://doi.org/10.4995/ILASS2017.2017.4846.

[30] I. Shopa, D. Jakubczyk, I. Kamińska, G. Derkachov, T. Wojciechowski, K. Fronc, Evaporation of levitating microdroplets containing nanoparticles: luminescent Gd2O3:Er3+, illumination with 806 nm laser from below via optical fiber, Mendeley Data (2024). https://doi.org/10.17632/jfdxfkhjy2.2.

[31] C.M. Sorensen, Light scattering by fractal aggregates: A review, Aerosol Sci. Technol. 35 (2001) 648–687. https://doi.org/10.1080/02786820117868.

[32] C.F. Bohren, D.R. Huffman, Absorption and Scattering of Light by Small Particles, Wiley, 1998. https://doi.org/10.1002/9783527618156.

[33] D. Jakubczyk, M. Kolwas, G. Derkachov, K. Kolwas, M. Zientara, Evaporation of microdroplets: The "radius-square-law" revisited, Acta Phys. Pol. A 122 (2012). https://doi.org/10.12693/APhysPolA.122.709.

[34] D. Tzarouchis, A. Sihvola, Light Scattering by a Dielectric Sphere: Perspectives on the Mie Resonances, Appl. Sci. 8 (2018) 184. https://doi.org/10.3390/app8020184.

[35] P. Chýlek, J.D. Pendleton, R.G. Pinnick, Internal and near-surface scattered field of a spherical particle at resonant conditions, Appl. Opt. 24 (1985) 3940. https://doi.org/10.1364/AO.24.003940.

[36] I. Machfuudzoh, T. Hinamoto, F.J. García de Abajo, H. Sugimoto, M. Fujii, T. Sannomiya, Visualizing the Nanoscopic Field Distribution of Whispering-Gallery Modes in a Dielectric Sphere by Cathodoluminescence, ACS Photonics 10 (2023) 1434–1445. https://doi.org/10.1021/acsphotonics.3c00041.

[37] C. Mätzler, MATLAB Functions for Mie Scattering and Absorption, IAP Res Rep 2002–08 (2002) 1139–1151. http://arrc.ou.edu/~rockee/NRA_2007_website/Mie-scattering-Matlab.pdf.

[38] S. Arnold, R. Ramjit, D. Keng, V. Kolchenko, I. Teraoka, MicroParticle photophysics illuminates viral bio-sensing, Faraday Discuss. 137 (2008) 65–83.

https://doi.org/10.1039/B702920A.

[39] L. Dong, D. Johnson, Surface Tension of Charge-Stabilized Colloidal Suspensions at the Water–Air Interface, Langmuir 19 (2003) 10205–10209. https://doi.org/10.1021/la035128j.

[40] D. Ngo, R.G. Pinnick, Suppression of scattering resonances in inhomogeneous microdroplets, J. Opt. Soc. Am. A 11 (1994) 1352. https://doi.org/10.1364/JOSAA.11.001352.

[41] G. Gobel, A. Lippek, T. Wriedt, K. Bauckhage, MONTE CARLO SIMULATION OF LIGHT SCATTERING BY INHOMOGENEOUS SPHERES, in: Proc. Second Int. Symp. Radiat. Transf., Begellhouse, Connecticut, 1997: pp. 1–10. https://doi.org/10.1615/ICHMT.1997.IntSymLiqTwoPhaseFlowTranspPhenCHTRadTransfProc.260.

[42] M. Kolwas, D. Jakubczyk, G. Derkachov, K. Kolwas, Interaction of optical Whispering Gallery Modes with the surface layer of evaporating droplet of suspension, J. Quant. Spectrosc. Radiat. Transf. 131 (2013) 138–145. https://doi.org/10.1016/j.jqsrt.2013.03.009.

[43] J. Archer, M. Kolwas, M. Woźniak, D. Jakubczyk, K. Kolwas, G. Derkachov, T. Wojciechowski, Sodium dodecyl sulfate microaggregates with diversely developed surfaces: Formation from free microdroplets of colloidal suspension, Eur. Phys. J. Plus 134 (2019) 39. https://doi.org/10.1140/epjp/i2019-12427-3.

[44] A.V. Lazo, P. Rathie, On the entropy of continuous probability distributions (Corresp.), IEEE Trans. Inf. Theory 24 (1978) 120–122. https://doi.org/10.1109/TIT.1978.1055832.

[45] G.C. Righini, Y. Dumeige, P. Féron, M. Ferrari, G.N. Conti, D. Ristic, S. Soria, Whispering Gallery Mode microresonators: Fundamentals and applications, Riv. Del Nuovo Cim. 34 (2011) 435–488. https://doi.org/10.1393/ncr/i2011-10067-2.

[46] J. Kher-Alden, S. Maayani, L.L. Martin, M. Douvidzon, L. Deych, T. Carmon, Microspheres with Atomic-Scale Tolerances Generate Hyperdegeneracy, Phys. Rev. X 10 (2020) 031049. https://doi.org/10.1103/PhysRevX.10.031049.